\documentclass[twocolumn,showpacs,aps]{revtex4}
\usepackage{graphicx}
\usepackage{epsfig}
\usepackage[american]{babel}
\usepackage{amsfonts} 
\usepackage[dvips]{color}

\usepackage[latin2]{inputenc}
\usepackage[T1]{fontenc}
\usepackage{graphics}
\usepackage{amsmath}
\usepackage{amsfonts}
\usepackage{amssymb}

\newcommand{\ReG}{\mbox{$Re$}}

\newcommand{\Sc}{Schr\"{o}dinger}

\begin{document}

\title{Quantum adiabatic polarons by translationally invariant perturbation theory}

\author{O. S. Bari\v si\' c$^1$ and S. Bari\v si\' c$^2$}      

\address{$^1$Institute of Physics, Bijeni\v cka c. 46, HR-10000 Zagreb, Croatia\\$^2$Department of Physics, Faculty of Science, University of
Zagreb, Bijeni\v cka c. 32, HR-10000 Zagreb, Croatia}

\begin{abstract}

The translationally invariant diagrammatic quantum perturbation theory (TPT) is applied to the polaron problem on the 1D lattice, modeled through the Holstein Hamiltonian with the phonon frequency $\omega_0$, the electron hopping $t$ and the electron-phonon coupling constant $g$. The self-energy diagrams of the fourth-order in $g$ are calculated exactly for an intermittently added electron, in addition to the previously known second-order term. The corresponding quadratic and quartic corrections to the polaron ground state energy become comparable at $t/\omega_0>1$ for $g/\omega_0\sim(t/\omega_0)^{ 1/4}$ when the electron self-trapping and translation become adiabatic. The corresponding non adiabatic/adiabatic crossover occurs while the polaron width is large, i.e. the lattice coarsening negligible. This result is extended to the range $(t/\omega_0)^ {1/2}>g/\omega_0>(t/\omega_0)^{1/4}>1$ by considering the scaling properties of the high-order self-energy diagrams. It is shown that the polaron ground state energy, its width and the effective mass agree with the results found traditionally from the broken symmetry side, kinematic corrections included. The Landau self trapping of the electron in the classic self-consistent, localized displacement potential, the restoration of the translational symmetry by the classic translational Goldstone mode and the quantization of the polaronic translational coordinate are thus all encompassed by a quantum theory which is translationally invariant from the outset. This represents the first example, open to various generalizations, of the capability of TPT to hold through the adiabatic symmetry breaking crossover. Plausible arguments are also given that TPT can describe the $g/\omega_0>(t/\omega_0)^ {1/2}$ regime of the small polaron with adiabatic or non-adiabatic translation, i.e., that TPT can cover the whole $g/\omega_0$, $t/\omega_0$  parameter space of the Holstein Hamiltonian.

\end{abstract}

\maketitle

\section{Introduction\label{SecIntroduction}}

The polaron is one of the earliest examples of a topological particle associated with symmetry breaking. Similar examples in the condensed physics are phasons \cite{F}, solitons \cite{Davydov}, magnetic polarons \cite{Gennes}, Kosterliz-Thouless vor\-ti\-ces \cite{KT}, Zhang-Rice singlets \cite{ZR}, and many others. Polaron was initially introduced by Landau \cite{L} as a state of an electron coupled to the classical deformation field, which, by adiabatic self-localization of the electron (electron self-trapping), breaks the original translational symmetry of the Hamiltonian. The corresponding symmetry restoring Goldstone mode (classical or quantum) is then the translation of the polaron \cite{LandauPekar,Shaw,Holstein4}.

This description of the (adiabatic) polaron, from the broken symmetry side, suggests that the translationally invariant quantum perturbation theory (TPT) in terms of the electron-phonon (e-p) coupling constant $g$, cannot reach such a (symmetry restored) state. Indeed, the symmetry breaking (at $T=0$) is, as a rule, associated with the singularity (quantum critical point, QCP) in the ground state energy as a function of $g$. This is usually taken to restrict the use of TPT to the high symmetry (translationally invariant) phase on one side of the QCP. On the other hand, the study of some continuous e-p Hamiltonians (including the 3D Fr\"ohlich Hamiltonian which exhibits a polaron as a classic solution), concluded that the corresponding quantum ground state energy is a smooth function of $g$ \cite{Gerlach}. This shows that, depending on the dimension of the system and the range of the forces, the quantum fluctuations of the deformation field can remove the QCP and replace it by a smooth crossover. However, this does not guarantee that TPT can hold through such a crossover \cite{A1,A2,A3}, because the absence of a singularity for $g$'s, which make the Hamiltonian hermitian, can say nothing about the radius of convergence and the behavior of the perturbation series in the complex $g$ plane. The question is not only whether TPT can hold through the symmetry breaking crossover in principle \cite{A4} but also how and for which physical regimes \cite{A5} this can be done in practice. Some elements of the answer to this question are given here.

In the adiabatic limit, which is of the main interest here, the electron is "self-trapped" (adiabatic self-trapping), staying always in the same localized state, which moves with the polaron distortion of the lattice. In the continuous approximation, the adiabatic polaron is free to move along the lattice. However, the lattice coarsening introduces the Peierls-Nabarro (PN) or Umklapp potential \cite{PN,Dz} into the motion of the adiabatic polaron. If treated classically, even a tiny periodic potential pins the polaron of mass $M_p$ to the lattice. There are thus two possible, separate, mechanisms which can break the translational symmetry of the lattice on the adiabatic level. The first is the adiabatic self-trapping of the electron. The second is the pinning of the adiabatic polaron to the PN potential of the lattice. However, although there are two symmetry breaking mechanisms, there is only one symmetry to break, namely the translational symmetry of the lattice. It is therefore important to distinguish between the electron adiabatic self-trapping and the polaron pinning. As the electron adiabatic self-trapping is prerequisite for the polaron pinning, it is the fundamental mechanism of symmetry breaking. For this reason, the problem of the applicability of the TPT is focused here on the electron adiabatic self-trapping on the lattice.

\begin{figure}[t]

\begin{center}{\scalebox{0.3}
{\includegraphics{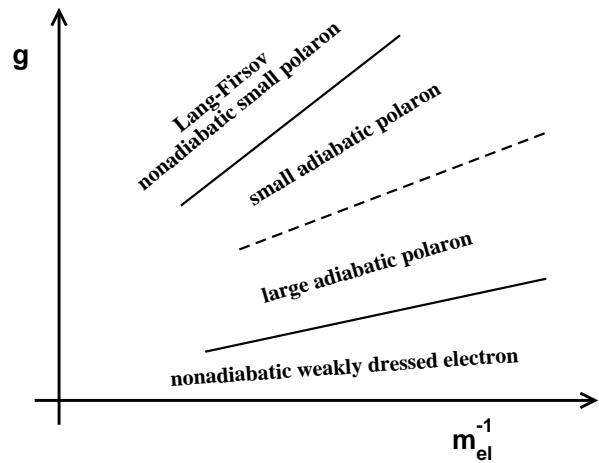}}}
\end{center}

\caption{Schematic phase diagram for the polaron problem in terms of the e-p coupling constant $g$ and the electron mass $m_{el}$ (arbitrary units). Full lines denote QCP's or crossovers, which could be responsible for the breaking of the translational symmetry of the lattice. The broken line denotes the crossover that is associated to the pinning of the polaron to the lattice.\label{Fig000}}

\end{figure}

In fact, the inter-relation of the electron self-trapping and the polaron pinning can be fully understood in the adiabatic regime. As illustrated schematically in Fig. \ref{Fig000}, this regime can be reached from the weak and from the strong coupling side. For small $g$, the quantum electron-phonon system is described reasonably well by the lowest order perturbation theory in the e-p coupling, which is apparently translationally invariant and nonadiabatic \cite{OSB}. For the discrete lattice the polaron motion is nonadiabatic also for extremely large $g$. The electron must use phonons nonadiabatically to leave its (too) strongly pinned adiabatic phonon correlation cloud in order to gain the delocalization energy. Obviously, this quantum state is also translationally invariant. The adiabatic polaron regime, if it exists, is thus separated from the two nonadiabatic, translationally invariant regimes either by a pair of QCP's or by a pair of corresponding crossovers. Which is the case can be found from either (small or large $g$) side. Again however, in the case when QCP's are replaced by crossovers, it remains to be proven that TPT can hold through them.

To make the picture complete, it should be further realized that on coming from the low $g$ side, the QCP or the crossover corresponding to the electron adiabatic self-trapping can, generally speaking, occur either towards the large adiabatic polaron quantum state, when the PN potential although present is negligible, or towards the small adiabatic polaron quantum state, when this potential plays an essential role. In the former case there is an additional crossover, the one from large to the small adiabatic polaron, as indicated by the dashed curve in Fig. \ref{Fig000}. Importantly, this behavior can never correspond to the breakdown of the translational symmetry of the lattice (i.e., to QCP), because, as already mentioned, the latter is consumed by the electron adiabatic self-trapping QCP or by the corresponding crossover. It appears therefore that the electron adiabatic self-trapping on the lattice provides a critical test for the applicability of TPT. In this context it is natural to study the low $g$ side of the problem, because TPT is the expansion in terms of $g$. If TPT is able to hold through this crossover or QCP it is likely that it applies, at least in principle, to all values of $g$. But then, so does also the theory which starts from the broken symmetry side, and it is only a matter of convenience which approach is to be used when.

In the present paper, TPT is tested on the discrete 1D Holstein model. This model provides a simple example of the general situation described above. For sufficiently fast electrons it exhibits a pair of nonadiabatic/adiabatic crossovers, which replace QCP's \cite{O}. The small $g$ crossover corresponds to the electron self-trapping in the state of the large adiabatic Holstein polaron \cite{Rashba,H}. At sufficiently large $g$ it is supplemented by the large/small adiabatic polaron crossover in the PN potential, which is thus well separated from the electron self-trapping crossover. It will be argued here that TPT can reach beyond the low $g$ electron self-trapping showing explicitly that this QCP is replaced by the crossover and that the lattice coarsening (PN, Umklapp) effects are negligible in this case. Finally, some plausible arguments are given which indicate that TPT is valid for all values of $g$, although the full proof of this assertion requires additional considerations.

\section{General}

The Holstein Hamiltonian on the discrete 1D lattice with $L$ sites is given by ($\hbar=1$)

\begin{equation}
\hat H=\sum_{-L/2}^{L/2-1}[-t\;c_r^\dagger(c_{r+1}+c_{r-1})+\omega_0\;b_r^\dagger b_r-g\;c_r^\dagger c_r(b_r^\dagger+ b_r)]
\label{Eq001}
\end{equation}

\noindent where the fermion and boson operators $c_r$ and $b_r$ on the site $r$ are defined in the usual way. Equation (\ref{Eq001}) is meant to describe $N$ electrons subjected to hopping $t$ along the chain and to the local interaction $g$ with the displacements of the lattice $u_r=x_0(b_r^\dagger+b_r)$. At $g=0$ the latter behave as harmonic oscillators with frequency $\omega_0$ and the zero-point displacement $x_0=\sqrt{1/2M\omega_0}$, where $M$ is the ionic mass. The properties of the Holstein Hamiltonian for a given $L$ and $N$ are thus described in the simple $2D$ parameter space, e.g. in terms of $g/\omega_0$ and $\alpha=\omega_0/2t$. The only adimensional quantity independent of $M$ that can be constructed from $g/\omega_0$ and $\alpha$ is $\Lambda=\alpha\;g^2/\omega_0^2$, which has to appear naturally (instead of $g^2/\omega_0^2$) in the adiabatic regime, of interest here.

The structure of TPT for the Hamiltonian (\ref{Eq001}) and similar models was examined in Refs. \cite{Fg,Mishchenko2,OSB}. The main conclusion in Ref. \cite{OSB} was that the polaron properties can be conveniently extracted from the correlation functions which form the basis of the diagrammatic version \cite{AGD} of TPT. Two types of correlation functions are to be distinguished in this respect, namely those which by construction conserve the number $N$ of electrons, and those which do not. The displacement-displacement correlation function $D$ belongs to the first class, and the electron propagator $G$ to the second, because it changes $N$ by $\pm1$. In the polaron case $D=D^{(1)}$ is to be taken at $N=1$ and, consistently, $G=G^{(0)}$ at $N=0$. $G^{(0)}$ describes the propagation of an electron intermittently added to the system of bosons. Actually, on using the Lehmann representation of $G^{(0)}$ it turned out \cite{OSB} that the most important properties of the polaronic correlations can be calculated from the corresponding irreducible electron self-energy $\Sigma^{(0)}$ alone. $\Sigma^{(0)}$ gives the position of the polaron bands, in particular the value of the ground state energy, the effective mass of the polaron, and, in the continuous limit, the polaron width, as will be further discussed below.

The reduction of the problem to the calculation of $\Sigma^{(0)}$ represents an important simplification, because at $N=0$ all fermionic (Pauli) correlations are eliminated from the outset. This contrasts with the calculation \cite{OSB} of the ground state energy from its direct diagrammatic expansion at $N=1$, where the electron exchange effects disappear from the result only after tedious cancellations.

\begin{figure}[tb]

\begin{center}{\scalebox{0.4}
{\includegraphics{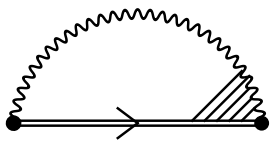}}}
\end{center}

\caption{The exact electron self-energy $\Sigma^{(0)}$. The double line is the exact $G$, the shaded triangle is the exact vertex and the wavy line is the free phonon propagator $D_0$.\label{Fig001}}

\end{figure}

This important simplification appears formally in the exact expression for $\Sigma^{(0)}$, given in Fig. \ref{Fig001} in the usual \cite{AGD} diagrammatic language. In contrast to the general $N$ case, the wavy line in Fig. \ref{Fig001} represents the free $N=0$ displacement-displacement correlation function $D^{(0)}=D_0$,

\[D_0=\frac{1}{2}\left[\frac{1}{\omega-\omega_0+i\eta}-\frac{1}{\omega+\omega_0-i\eta}\right]\;.\]

\noindent Indeed, at $N=0$, the phonon renormalization, which starts necessarily with the creation of the electron-hole pair, is impossible: there is no electron in the system, additional to that created intermittently. In other words, the electron described by $G^{(0)}$ can only advance in time in Fig. \ref{Fig001} until it is annihilated, i.e., $G^{(0)}$ can have poles only in the lower $\omega$-half-plane. This holds in particular for the free electron propagator $G_0^{(0)}$,

\begin{equation}
G_0^{(0)}(k,\omega)=\frac{1}{\omega-\xi_k+i\eta}\;.\label{Eq003}
\end{equation}

\noindent For further convenience the zero of the free electron energy $\xi_k$ in Eq. (\ref{Eq003}) is taken, unlike in Eq. (\ref{Eq001}), at $k=0$,

\begin{equation}
\xi_k=2t(1-\cos{(k)})\;.\label{Eq004}
\end{equation}

\begin{figure}[tb]

\begin{center}{\scalebox{0.5}
{\includegraphics{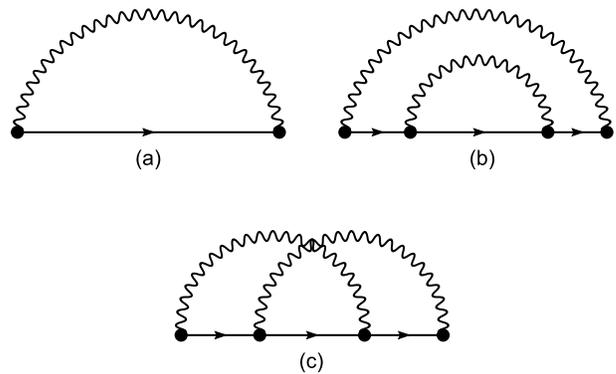}}}
\end{center}

\caption{Second- and fourth-order diagrams for $\Sigma^{(0)}$.\label{Fig002}}

\end{figure}

Figure \ref{Fig001} also contains the triangular vertex correction of Fig. \ref{Fig002}c originally discussed by Migdal \cite{Mi} in the limit of a large number of electrons $N$. At large $N$, especially when the soft-phonon renormalizations of the phonon propagator are important, this vertex correction can sometimes be neglected. However, at $N=0$, when $D^{(0)}=D_0$ in Fig. \ref{Fig001}, the vertex correction can be as important as the corresponding self-energy correction, as will be further seen below. The fact that the "soft-phonon" corrections do not appear at $N=0$ in Fig. \ref{Fig001} does not mean that the phonon propagator $D^{(1)}$ at $N=1$ is not developing a soft-phonon branch, as a signature of the polaronic correlations. It does, but this branch does not enter the calculations of $\Sigma^{(0)}$ in the small-$N$ hierarchy of the correlation functions.

In order to emphasize that at $N=0$ the electron can only advance in time, the irreducible $\Sigma^{(0)}$ of Fig.\ \ref{Fig001} can be represented by the expansion, order by order in $g$. In Fig. \ref{Fig002} the arrow on the free electron line $G^{(0)}$ of Eq. (\ref{Eq003}) points explicitly forwards in time. As usually, the diagrams in Fig. \ref{Fig002} represent the perturbation series in $g$ either in the direct or in the reciprocal space. Concerning the propagation in space it should be noted that $\Sigma_2$ and the non-crossing contribution $\Sigma_4^{NC}$ are local because so is $D_0$ of Eq. (\ref{Eq004}), in contrast to the crossing diagram $\Sigma_4^C$ which contains also the phonon-assisted intersite propagation. In terms of Fig. \ref{Fig001}, $\Sigma_4^C$ represents the leading (Migdal) vertex correction to $\Sigma^{(0)}$. It will be shown below that in the large adiabatic polaron regime the non-local contributions of the crossing diagrams to the low-energy properties are equally important for the polaron formation as the local corrections of the NC diagrams.

The low-energy properties are associated with the polaron spectrum $\tilde\xi_k$. Assuming that the perturbation series for $\Sigma^{(0)}$ shown in Fig. \ref{Fig002} is meaningful, $\tilde\xi_k$ is the solution of the equation

\begin{equation}
\tilde\xi_k=\xi_k+\ReG\Sigma^{(0)}(\xi_k,\tilde\xi_k)\;,\label{Eq005}
\end{equation}

\noindent where $\ReG\Sigma^{(0)}$ represents the real part of $\Sigma^{(0)}$. It is implicit in the concept of the quantum polaron, as of the coherent quantum state, that the imaginary part of $\Sigma^{(0)}$ is infinitesimally small around the polaron pole $\tilde\xi_k$. This is shown later to be consistent with the present analysis and, anticipating this result, no distinction is made further between $\ReG\Sigma^{(0)}$ and $\Sigma^{(0)}$ itself. The energy of the $k=0$ state in the lowest polaron band is obviously related to the ground state energy. Noting that the ground state energies $E_0^{(1)}(0)$ and $E_0^{(0)}(0)$ of the noninteracting ($g=0$) $N=1$ system and the $N=0$ system, respectively, are taken to coincide by choosing $\xi_{k=0}=0$ in Eq. (\ref{Eq004}), the ground state energy gain $\Delta$ of the $N=1$ system with finite $g$ is

\begin{equation}
\Delta=E_0^{(1)}(0)-E_0^{(1)}(g)=-\Sigma^{(0)}(k=0,-\Delta)\;,\label{Eq006}
\end{equation}

\noindent according to the Lehmann representation of $G^{(0)}$. It is worthy of note that the higher energy solutions of Eq. (\ref{Eq006}) give the values of the polaron $k=0$ energies of the excited polaron bands. It is reiterated that the upper index in Eq.~(\ref{Eq006}) and elsewhere denotes the number of electrons involved.

The behavior of $\tilde\xi_k$ in the vicinity of $k=0$ defines the polaron mass $M_p$ which is given by the derivates of $\Sigma^{(0)}(\xi_k,\omega)$ at $k=0$ and $\omega=-\Delta$,

\begin{equation}
M_p^{-1}=2t\;\frac{1+\partial\Sigma^{(0)}/\partial\xi_k}
{1-\partial\Sigma^{(0)}/\partial\omega}\;.\label{Eq007}
\end{equation}

\noindent It is worth noting that the denominator in Eq. (\ref{Eq007}) taken at arbitrary $k$ gives the corresponding electron spectral density.

Finally, it will be shown latter that the polaron width can be related to $\Delta$ of Eq. (\ref{Eq006}), in particular for the large adiabatic polaron. This completes the list of the main polaron properties (energy, mass, width) which can be derived from $\Sigma^{(0)}(k,\omega)$. The next step is the examination of $\Sigma^{(0)}(k,\omega)$ itself.

\section{Low-order diagrams}

The behavior of the leading (skeleton diagram) diagram $\Sigma_2$ (dropping henceforth the $(0)$ superscript) of Fig. \ref{Fig002} is quite instructive \cite{OSB}. $\Sigma_2$ is given as

\begin{equation}
\Sigma_2=-\frac{g^2}{4\pi t}I(\varepsilon)\label{Eq008}
\end{equation}

\noindent with

\begin{equation}
I(\varepsilon)=\frac{2\pi}{L}\sum^{L/2-1}_{-L/2}\frac{1}{\varepsilon+2\sin^2{(\pi m/L)}}\;,\label{Eq009}
\end{equation}

\noindent where

\begin{equation}
\varepsilon=\frac{\omega_0-\omega}{2t}=\alpha-\frac{\omega}{2t}\;,\label{Eq010}
\end{equation}

\noindent and the integer $m$ is related to the phonon wavevector $q$ by the Born-van K\` arm\`an boundary conditions,

\begin{equation}
q=2\pi m/L\;.\label{Eq011}
\end{equation}

\noindent The sum (\ref{Eq009}) can be expressed in the closed, analytical form for arbitrary $\varepsilon>0$. The regime of interest here is $L^{-2}<\varepsilon\ll1$. For $\varepsilon>L^{-2}$ finite size effects can be neglected, i.e., the sum in Eq. (\ref{Eq009}) can be turned into the integral over $q$. If $\varepsilon\ll1$, $q_c=\sqrt{2\varepsilon}$ plays the role of the infrared cutoff in this integration. In other words, $2\sin^2{(q/2)}=1-\cos(q)$ in Eq. (\ref{Eq009}) can be replaced then by $q^2/2$ and integrated from $q_c$ to infinity, yielding

\begin{equation}
I(\varepsilon)\approx \pi\sqrt{2/\varepsilon}\;.\label{Eq012}
\end{equation}

\noindent The same result is obtained by taking the continuous limit in the Hamiltonian (\ref{Eq001}) from the outset, except that the present procedure shows that the corrections due to the lattice coarsening (to the Umklapp processes) are of the order of $\varepsilon\ll1$ itself.

In the weak-coupling limit, when the ground state energy shift $\Delta$ of Eq. (\ref{Eq006}) is small $\Delta\ll\omega_0$, $\omega=-\Delta$ can be neglected in $\varepsilon$ of Eq. (\ref{Eq010}) appearing in combined equations (\ref{Eq006}) and (\ref{Eq012}). Taking thus $\varepsilon\approx\alpha\ll1$ leads to

\begin{equation}
\Delta\approx\frac{g^2}{\omega_0}\sqrt\alpha=2t\Lambda\sqrt\alpha \;.\label{Eq013}
\end{equation}

\noindent This apparently is a nonadiabatic result, because $\Delta$ is dependent on the ionic mass $M$ through $\alpha$. In the weak-coupling limit $L^{-2}<\varepsilon\approx\alpha\ll1$ thus ensures validity of the continuous \cite{OSB} rather than of the adiabatic \cite{Fg} approximation. Alternatively, the inequality $L^{-2}<\varepsilon\ll1$ may be thought \cite{OSB} of as fixing the order of the limits, first $L\rightarrow\infty$, and then $t/\omega_0\rightarrow\infty$, in the search for a nontrivial adiabatic regime. 

In this context, the nonadiabatic nature of Eq. (\ref{Eq013}) for $\alpha<1$ suggests the calculation of the higher order diagrams $\Sigma_4^{NC}$ and $\Sigma_4^C$ of Fig. \ref{Fig002} in the first place. The two internal frequency integrations in those diagrams can be carried out easily to give

\begin{eqnarray}
\Sigma_4^{NC}&=&\frac{g^4}{(2t)^3}\;I'(\varepsilon)\;I(\varepsilon+\alpha) \label{Eq014}\\
\Sigma_4^C&=&-\frac{g^4}{(2t)^3}\sum_{q,q'}
\frac{1}{[\varepsilon+2\sin^2{(q/2)}]} \frac{1}{[\varepsilon+2\sin^2{(q'/2)}]}\nonumber\\
&\times&
\frac{1}{[\varepsilon+\alpha+2\sin^2{((k+q-q')/2)}]}\;.\label{Eq015}
\end{eqnarray}

\noindent where $I(\varepsilon)$ in Eq. (\ref{Eq014}) is given by Eq. (\ref{Eq009}), $I'(\varepsilon)$ is its derivative with respect to $\varepsilon$, while $q,q'$ and $k$ in Eq. (\ref{Eq015}) are given by Eq. (\ref{Eq011}). Translational invariance was used in Eq. (\ref{Eq015}) to associate the dependence of $\Sigma_4^C$ on the external momentum $k$ with the convolution of $q$ and $q'$, i.e., with the crossing of two phonon lines in Fig. \ref{Fig002}c. In contrast, as already mentioned, $\Sigma_4^{NC}$ (like $\Sigma_2$) is local, i.e., independent of $k$.

While $I(\varepsilon)$ and thus $I'(\varepsilon)$ of Eq. (\ref{Eq014}) are known in the closed form for $\varepsilon>0$ arbitrary, $\Sigma_4^C$ calculated in Appendix is exhibited here only in the interesting limit $L^{-2}<\varepsilon+\alpha\ll1$, when the $q,q'$ summations in Eq. (\ref{Eq015}) can be turned into the infrared singular integrations. Together with Eqs. (\ref{Eq014}) and (\ref{Eq012}) this gives

\begin{eqnarray}
\Sigma_4^{NC}&=&-\frac{g^4}{(2t)^3}\frac{1}
{\sqrt{\varepsilon^3(\varepsilon+\alpha)}}\label{Eq016}\;,\\
\Sigma_4^C&=&-\frac{g^4}{(2t)^3}
\frac{2\sqrt\varepsilon+\sqrt{\varepsilon+\alpha}}
{\varepsilon\sqrt{\varepsilon+\alpha}}
\frac{1}{k^2+2(2\sqrt\varepsilon+\sqrt{\varepsilon+\alpha})^2}\;,\label{Eq017}
\end{eqnarray}

\noindent both results holding with the accuracy $1/(\varepsilon+\alpha)$.

Equations (\ref{Eq016}) and (\ref{Eq017}) can be used first to define the range of validity of the low-order perturbation theory by comparing $\Sigma_2$ and $\Sigma_4$ at $k=0$ and $\omega=-\Delta$ of Eq.~(\ref{Eq006}). The two become comparable for $\Delta$ comparable to $\omega_0$, i.e., for

\begin{equation}
g/\omega_0\approx\alpha^{-1/4}\;\;\;\Rightarrow\;\;\;\Lambda\alpha^{-1/2}\approx1\;,\label{Eq018}
\end{equation}

\noindent with $\alpha,\Lambda\ll1$. In Ref.\cite{OSB} the crossover condition (\ref{Eq018}) was derived from $D^{(1)}$ (at $N=1$), on considering the average number of excited phonons and on requiring that the latter is at most of the order of unity. Here, this condition follows from the electron self-energy $\Sigma^{(0)}$, on the same physical grounds, determining when the two-phonon processes in Fig. \ref{Fig002} are becoming equally important as the single-phonon processes.

The main contribution to $\Sigma_4^C$ comes from the non-local ($k<\varepsilon+\alpha$) phonon-assisted processes. Indeed, the local contribution of $\Sigma^C_4$, obtained by integrating Eq.~(\ref{App02}) or approximately the long-wave limit of this expression (\ref{App03}) over $k$, is apparently negligible with respect to the local $\Sigma_4^{NC}$. In other words, for $\varepsilon+\alpha\ll1$ the local contribution of the quartic non-crossing diagram and the non-local $k\approx0$ contribution of the quartic diagram are equally important for the values of parameters satisfying Eq.~(\ref{Eq008}) in determining the ground state energy and the polaron mass, given by Eqs. (\ref{Eq006}) and (\ref{Eq007}), respectively. This regime is thus beyond the reach of the DMFT \cite{Fg}.

As has been already emphasized before, the condition (\ref{Eq018}) when introduced in Eq. (\ref{Eq013}) leads to

\begin{equation}
\Delta\approx2t\Lambda^2\;,\label{Eq019}
\end{equation}

\noindent for $\alpha\ll1$, where, once again, $\Lambda=g^2/2t\omega_0$ is independent of ionic mass. The estimate (\ref{Eq019}) for the ground state energy is therefore adiabatic, i.e., the condition (\ref{Eq018}) corresponds to the nonadiabatic/adiabatic crossover line in the $g/\omega_0,\alpha$ parameter space of the Holstein Hamiltonian. Actually, the result (\ref{Eq019}) is, up to the numerical coefficient, the same as the obtained from the symmetry broken side \cite{Emin,Holstein4}, by the self-trapping of the electron in the continuous version of the Holstein Hamiltonian (\ref{Eq001}). In this approach, the condition (\ref{Eq018}) appears as restricting the continuous adiabatic theory to the values $\Lambda\alpha^{-1/2}>1$, $\Lambda<1$. For $\Lambda\alpha^{-1/2}\approx1$ the nonadiabatic corrections are becoming appreciable (resulting finally in Eq. (\ref{Eq013}) for small couplings). On the other hand, associating $\Lambda^{-1}$ through $\Delta=g^2/\omega_0d$ with the polaron width $d$ by $d\sim\Lambda^{-1}$, the condition $\Lambda\ll1$ keeps the continuous theory valid not only on the crossover line $\Lambda\alpha^{-1/2}\approx1$ but also for $\Lambda\alpha^{-1/2}\gg1$.

It remains thus to be shown within TPT that the infrared singular (continuous), adiabatic result (\ref{Eq019}) holds not only on the crossover line (\ref{Eq018}) in the 2D parameter space but also for $\Lambda\alpha^{-1/2}\gg1$, as long as the continuous (infrared singular) approximation is valid due to $\Lambda\ll1$. Such a step amounts to the demonstration that TPT can reach beyond the nonadiabatic/adiabatic crossover (\ref{Eq018}), associated with the symmetry breaking. Apparently this requires the consideration of the infinite TPT series in Fig. \ref{Fig002}.

\section{Infinite series}

\begin{figure}[thb]

\begin{center}{\scalebox{0.6}
{\includegraphics{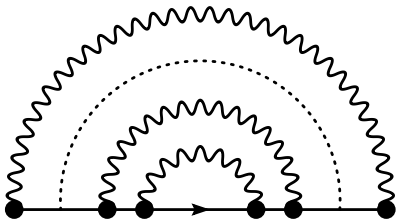}}}
\end{center}

\caption{The class of non-crossing diagrams corresponding to Eq. (\ref{Eq022}).\label{Fig005}}

\end{figure}

The general diagram of Fig. \ref{Fig002} can be evaluated in principle by using the usual diagrammatic sum rules \cite{AGD}. The class of non-crossing (NC) diagrams shown in Fig. \ref{Fig005}, obtained by generalizing $\Sigma_4^{NC}$ to the order $2p$, are especially simple because they are local. The structure of the $p$th order NC diagram, proportional to $g^{2p}$, is easily understood by noting that the external electron phonon bubble differs from $\Sigma_2$ by the fact that the single $G_0$ in $\Sigma_2$ is cut in two $G_0$'s by the ($p-1$) insert. Cutting $G_0$ in two amounts to the taking derivative of $\Sigma_2$ with respect to a parameter in $G_0$, analogously to the text-book demonstration of the Ward identities. For $p=2$ this procedure gives immediately the result (\ref{Eq014}) and iterates it to the order $p$, with $I(\varepsilon)$ given by Eq. (\ref{Eq009}),

\begin{equation}
\Sigma_{2p}^{NC}=A_p^{NC}\frac{g^{2p}}{t^{2p-1}}
I'(\varepsilon)\;I'(\varepsilon+\alpha)\;...\;
I(\varepsilon+(p-1)\alpha)\;.\label{Eq022}
\end{equation}

\noindent The numerical coefficients $A_p^{NC}$ are not of particular interest here, as only the scaling properties of $\Sigma$ are considered.

\begin{figure}[thb]

\begin{center}{\scalebox{0.6}
{\includegraphics{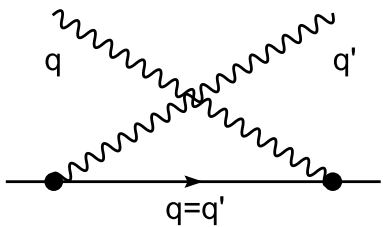}}}
\end{center}

\caption{Decoupling of infrared integrations over $q$ and $q'$ in Eq. (\ref{Eq017}).\label{Fig003}}

\end{figure}

The situation is more intricate with crossing diagrams, except in the particularly simple limit $t\approx0$ ($\alpha\approx\infty$), where the result has been obtained to all order in $g$. In order to make the discussion more transparent, $k$ is set equal to zero (as appropriate for the calculation of the ground state energy). In addition, unlike in Eq.~(\ref{Eq022}) or Eq.~(\ref{App02}), but as in Eq.~(\ref{Eq017}), the infrared limit is taken from the outset on assuming

\begin{equation}
\varepsilon+(p-1)\;\alpha<1\;,\label{Eq023}
\end{equation}

\noindent i.e., $\varepsilon$ and $\alpha$ small.

The $p=2$ example of this procedure is given by Eq. (\ref{Eq017}) taken at $k=0$. It is instructive to study first this example further in the limit $\varepsilon\gg\alpha$ (keeping $\varepsilon+\alpha\ll1$), anticipating that, for $g/\omega_0>\alpha^{-\frac{1}{4}}$ the physically relevant value of $\varepsilon$, $\varepsilon_\Delta$, is much larger than $\alpha$, i.e., that $\Delta\gg\omega_0$. In this limit both contributions $\Sigma_4^{NC}$ and $\Sigma_4^C$ of Eqs.~(\ref{Eq016}) and (\ref{Eq017}) have the same leading behavior in $\varepsilon$ large, $\Sigma_4\sim g^4/t^3\varepsilon^2$. This result can be derived simply not only for $\Sigma_4^{NC}$ but for $\Sigma_4^C$ too, on noting that the main contribution to the infrared singularity in the limit (\ref{Eq023}) with $\varepsilon\gg\alpha$ comes from $q\approx q'$ in Eq.~(\ref{Eq015}). Setting $q=q'$ in the convolution term, the infrared integrations over $q$ and $q'$ decouple, each yielding the contribution given by Eq.~(\ref{Eq012}), the overall results being $\Sigma^C_4\sim g^4/t^3\varepsilon^2$. However, the described approximation affects the numerical prefactor and the $\alpha/\varepsilon$ corrections in $\Sigma_4^C$, which are thus out of control in this approximation. In return, the benefit of described procedure is that it is apparently iterative in the sense that in the general $p$th order diagram it applies, under the condition (\ref{Eq023}), to any segment of the type shown in Fig. \ref{Fig003}. This suggests that the full $k=0$ irreducible $p$th order self-energy contribution $\Sigma_{2p}$ has the same form as $\Sigma_{2p}^{NC}$ of Eq. (\ref{Eq022}) in the infrared limit, i.e.,

\begin{equation}
\Sigma_{2p}=2t\;A_p\left(\frac{g}{2t}\right)^{2p}\;
\frac{1}{\varepsilon^{\frac{3p}{2}-1}}\;\sigma_{2p}(\varepsilon/\alpha)\;,\label{Eq024}
\end{equation}

\noindent with the numerical coefficients $A_p$ and the function $\sigma_{2p}$ undetermined here. Importantly, however, for later discussion, $\sigma_{2p}(\varepsilon/\alpha)$ tends to a finite constant for $\varepsilon/\alpha$ large provided that $p$ satisfies the condition (\ref{Eq023}). All diagramms of the sixth-order in $g$ checked satisfy Eq. (\ref{Eq024}).

Turning to the scaling properties of $\Sigma_{2p}$ it should be noted that, due to its particular structure in $\alpha$, the series (\ref{Eq024}) can be rescaled from the scale $2t$ to the scale $\omega_0$, by introducing

\[\eta=\varepsilon/\alpha=\frac{\omega_0-\omega}{\omega_0}\;,\]

\noindent to obtain 

\begin{eqnarray}
\Sigma_{2p}&=&\omega_0\;A_p\;\sigma_{2p}(\eta)\;
(\Lambda\alpha^{-1/2})^p/\eta^{\frac{3p}{2}-1}\;.\label{Eq025}
\end{eqnarray}

\noindent This scaling of $\Sigma_{2p}$ has many satisfactory features. $\Sigma_{2p}$ is proportional to $\omega_0$ for any $p$, showing explicitly that not only the crossing diagrams at $k\approx0$ are proportional to $\omega_0$ ("Migdal theorem") but that so are the NC ones, and thereby that both have to be treated on equal footing. In fact, the number of crossing diagrams $n^C$ is large for large $p$ ($n^C=1$ for $p=2$, $n^C=8$ for $p=3$). Even more fundamentally, the proportionality to $\omega_0$ shows that the quantum polaron is being considered here, because $\omega_0$ in Eq. (\ref{Eq025}) is to be replaced everywhere by $\hbar\omega_0$ when $\hbar$ is not taken equal to $1$ in Eq.~(\ref{Eq001}). Equation~(\ref{Eq025}) also shows that $\omega_0$ should not to be set equal to zero too early \cite{Fg} (or inconsistently) when the adiabatic limit is considered. In this respect it is also gratifying that the adiabatic parameter $\Lambda$ is exhibited explicitly in Eq.~(\ref{Eq025}).

Equation (\ref{Eq025}) is also handy in the sense that, after using $\varepsilon\ll1$, $\alpha\ll1$ in the inequality (\ref{Eq023}), to produce the continuous limit in Eq. (\ref{Eq024}), it can use (the physical values) $\eta=\varepsilon/\alpha\gg1$ to make the prefactor of $(\Lambda\alpha^{-1/2})^p$ small when $\Lambda\alpha^{-1/2}$ is large. This provides a strong argument in favor of the convergence of the series (\ref{Eq025}), although the coefficients $A_p$ and $\sigma_{2p}(\infty)$ are unknown. Here, it can be only noted that Eq. (\ref{Eq025}) is valid for arbitrary large $p$ in the limit $\alpha\ll1$ provided that the physical relevant values of $\varepsilon$ are sufficiently small, $\varepsilon\ll1$.

\section{Quantum polaron properties from TPT}

Assuming thus that the infinite series (\ref{Eq025}) is convergent in the sense that it defines a function 

\[\Sigma=2t\;\alpha\;F(\eta,\;\Lambda\alpha^{-1/2})\;,\]

\noindent the physically relevant values $\eta_\Delta$ of $\eta$ are defined by Eq. (\ref{Eq005}). In particular, Eq. (\ref{Eq006}) for $\Delta$ can be conveniently rewritten as

\begin{equation}
\eta_\Delta-1=F(\eta_\Delta,\;\Lambda\alpha^{-1/2})\;,\label{Eq026}
\end{equation}

\noindent using Eq. (\ref{Eq010}). This shows that $\eta_\Delta $ can be expressed as

\begin{equation}
\eta_\Delta-1=f(\Lambda\alpha^{-1/2})\;.\label{Eq027}
\end{equation}

\noindent Noteworthy is the fact that when $\Sigma$ is real, as it turns out to be in Eq.~(\ref{Eq025}), the diagrammatic TPT of Eq. (\ref{Eq026}) is equivalent to the Wigner TPT. The \Sc\ TPT leads straightforwardly to Eq. (\ref{Eq027}) but through a series for which the convergence properties are less transparent than those of Eq. (\ref{Eq025}). The scaling relations of the type (\ref{Eq027}) are common in infrared problems \cite{U}.

The function $f$ in Eq. (\ref{Eq027}) has apparently two regimes $\Lambda\alpha^{-1/2}\lessgtr1$ with the crossover at $\Lambda\alpha^{-1/2}\approx1$. For $\Lambda\alpha^{-1/2}<1$ it can be expanded as a Taylor series in $\Lambda\alpha^{-1/2}$ with a constant term omitted. Keeping in mind that $\Delta$ vanishes at $g=0$ this immediately gives the nonadiabatic result (\ref{Eq013}), to the lowest order in $g$. The crossover at $\Lambda\alpha^{-1/2}\approx1$ is also already discussed in Eqs. (\ref{Eq018}) and (\ref{Eq019}), as occurring towards the large, adiabatic Holstein polaron. The true test of the ability of TPT to reach beyond this crossover corresponds, as already mentioned, to the regime $\Lambda\alpha^{-1/2}>1$ for $\Lambda\ll1$.

The only way for Eq. (\ref{Eq027}) to have the adiabatic solution for $\Lambda\alpha^{-1/2}\gg1$ is that the corresponding asymptotical behavior of $f(\Lambda\alpha^{-1/2})$ is

\begin{equation}
f(\Lambda\alpha^{-1/2})=\Lambda^2\alpha^{-1}
[c_0+c_1\alpha^{1/2}\Lambda^{-1}+c_2\alpha\Lambda^{-2}+...]\;,\label{Eq028}
\end{equation}

\noindent because only then $\alpha$, i.e., $\omega_0^{-1}\sim\sqrt M$, cancels out (rather than goes to zero) from Eq. (\ref{Eq027}), to the leading order in $\Lambda^{-1}\alpha^{1/2}$ small. The final result is then

\begin{equation}
\Delta=2t\Lambda^2
[c_0+c_1\alpha^{1/2}\Lambda^{-1}+c_2\alpha\Lambda^{-2}+...]\;.\label{Eq029}
\end{equation}

\noindent The result (\ref{Eq029}) satisfies the initial assumptions $\alpha\ll1$, $\varepsilon_\Delta\ll1$, $\eta_\Delta=\varepsilon_\Delta/\alpha\gg1$, which led to Eqs. (\ref{Eq024}-\ref{Eq026}). Two other equivalent ways to express those conditions are $\alpha\ll1$, $\Lambda\ll1$, $\Lambda\alpha^{-1/2}\gg1$, or $\omega_0/2t\ll1$, $\Delta/2t\ll1$, $\Delta/\omega_0\gg1$.

The result (\ref{Eq029}) not only reproduces the ground state energy (\ref{Eq019}) \cite{Rashba,H}, but also agrees with its corrections \cite{Holstein4}. The evaluation of these from the broken symmetry side is quite intricate, providing the first few coefficients: the even coefficients $c_0$, $c_2$, $c_4$, are finite and the odd ones, $c_1$ and $c_5$, vanish. The coefficient $c_2$ is associated with the energy proportional to $\omega_0$ and independent of $\Lambda$. This energy corresponds to the adiabatic reduction of the zero-point energy of the local harmonic modes which describe the polaron deformation. $c_4$ corresponds to the nonadiabatic kinematic corrections. The vanishing of the odd coefficients can be understood upon noting that when the leading term for $\Delta$ in Eq. (\ref{Eq029}) is inserted in $\eta_\Delta$ of Eq. (\ref{Eq025}), 

\[\Sigma_{2p}^{(0)}\sim\left(\frac{1}{\Lambda^2\alpha^{-1}}\right)^{p-1}\;,\]

\noindent i.e., $f$ of Eq. (\ref{Eq027}) becomes a function of $\Lambda^{-2}\alpha$ rather then of $\Lambda\alpha^{-1/2}$.

It can be objected that Eq. (\ref{Eq029}) follows from the assumption that TPT is convergent and adiabatic, rather than from the corresponding rigorous proofs. However, TPT is not worse in this sense than the broken symmetry approach. The latter also starts by assuming that the large adiabatic polaron exists and subsequently derives its properties and the conditions of validity. It is therefore all the more gratifying that those properties appear to be the same from the translationally invariant and from the broken symmetry side.

Actually, it is possible to extend the agreement between the two approaches from the condensation energy (\ref{Eq029}) to the polaron with $d$ and to the polaron mass $M_p$. In the continuous limit no ground state energy is associated with the polaron motion, i.e., the whole $\Delta$ is associated with the electron localization and with the concomitant displacements within the polaron width $d$. The localization energy to one site is $g^2/\omega_0$ and representing $\Delta$ as $\Delta=g^2/\omega_0\;d$, it follows that

\[d=\Lambda^{-1}[c_1'+c_2'\alpha^{1/2}\Lambda^{-1}+...]\;.\]

\noindent The kinematic corrections to the main Holstein result $d\sim\Lambda^{-1}$ have not been considered before. A more rigorous proof of the relation $\Delta=g^2/\omega_0\;d$ requires the consideration of spatial correlations, the task in which the diagrammatic TPT has certain technical advantage over the Wigner or \Sc\ TPT.

The same type of scaling can be applied to the finite $k$ states in Eq. (\ref{Eq005}) and in particular to the effective polaron mass (\ref{Eq007}). The derivate $\partial\Sigma^{(0)}/\partial\omega$ at $\omega=-\Delta$, associated with the electron spectral density, is given by the function $F(\eta,\;\Lambda\alpha^{-1/2})$ of Eq. (\ref{Eq026}). On the other hand the evaluation of $\partial\Sigma/\partial\xi_k$ requires the generalization of Eq.~(\ref{Eq017}), taken at finite $k$, to all crossing diagrams. Instead of that it is convincing enough to note that Eq.~(\ref{Eq017}) makes it plausible to set

\[\frac{1}{2t\;M_p}=h(\Lambda\alpha^{-1/2})\;,\]

\noindent in analogy with Eq. (\ref{Eq027}). The asymptotic behavior of the function $h$ for $\Lambda\alpha^{-1/2}$ large must give $M_p$ linear in $M\sim\alpha^{-2}$ to the leading order in $\Lambda^{-1}\alpha^{1/2}$ small, i.e., $M_p\sim g^8$. This also is a well-known result, derived previously from the broken symmetry side. On the other hand, the scaling properties of the electron spectral density in $M$, and then in $g$, or vice-versa, remain to be determined because the present scaling approach only connects the two.

\section{Concluding remarks}

The central conclusion of this paper is that the properties of the large adiabatic Holstein polaron on the 1D lattice can be determined by the TPT summed to the infinite order in $g$. The main properties, such as the polaron condensation energy, its width and the effective mass, can all be determined from the self-energy associated with the intermittent addition of one electron to the system of bosons.

Actually, a scaling analysis rather than a precise calculation was carried out under the assumption that the infinite order TPT series defines a function, which has a meaningful asymptotic behavior in the strong-coupling limit: this is termed a "convergence" of TPT. The analysis in question was applied to the extreme limit $\Lambda\alpha^{-1/2}\gg1$ and $\Lambda\ll1$, $(\alpha\ll1$), which ensure the validity of, respectively, the adiabatic and the continuous approximation. Although a full agreement is obtained with the corresponding results derived from the symmetry broken side it is of interest, as explained in Sec. \ref{SecIntroduction}, to supplement the discussion with a few remarks which concern the applicability of TPT to other cases.

The inequality (\ref{Eq023}), taken in the ground state, i.e., with $\eta=\eta_\Delta$ of Eq.~(\ref{Eq026}) defines the highest order $\Sigma_{2\bar p}$ of Eqs. (\ref{Eq024}) and (\ref{Eq025}) for which $\Delta\gtrsim\bar p\;\omega_0$. It is obvious that $\bar p$ is the average number of phonons in the system excited by the added electron for a given $\Lambda\alpha^{-1/2}$. In Eq. (\ref{Eq026}) $\bar p$ is taken to infinity, but it is interesting to consider the situation when $\bar p$, i.e., $\Lambda\alpha^{-1/2}$ is large but finite. For $p\lesssim\bar p$ the self-energy $\Sigma_{2p}$ is infrared singular and, according to Eq. (\ref{Eq025}), $\Sigma_{2p}\sim(\eta_\Delta)^{1-3p/2}$ to the leading order in $\eta$ large. For $\bar p<p<\alpha^{-1}$ the self-energy $\Sigma_{2p}$ is still infrared singular, $\Sigma_{2p}\sim(\eta_\Delta)^{-3\bar p/2}(p/\bar p)^{-1/2}$. However, for $p\gg\alpha^{-1}$ (finite) $\Sigma_{2p}$ tends to the well-known \cite{Fg} discrete, nonadiabatic $t\approx0$, small bandwidth limit $\Sigma_{2p}\sim(1/p!)^2$.

These observations help to elucidate the nature of the corrections to the large adiabatic polaron, additional to the non-adiabatic $c_2$, $c_4$ corrections, present already in the continuous limit. It appears that the Umklapp corrections to the infrared $\Sigma_{2p}$ for $p\lesssim\bar p$ are presumably responsible for the build up of the PN barrier. Indeed, the latter are known to become important for $\bar p\approx\alpha^{-1}$ (i.e., for $\Lambda\approx1$), when the other candidate for Umklapp corrections, namely the infrared terms $\Sigma_{2p}$ with $\bar p<p<\alpha^{-1}$, are squeezed out. On the other hand, the terms at $p>\alpha^{-1}$ are expected to lead to the non-adiabatic corrections to the translational dynamics of the small polaron. The strength of the $p\lesssim\bar p$ terms can be evaluated using a scaling procedure different from Eq.~(\ref{Eq028}), which shows that the PN potential is adiabatic, i.e., depends only on $t$ and $\Lambda$, and is exponentially small \cite{Kivshar} for $\Lambda\ll1$.

This reasoning shows that the TPT series evolves smoothly from one regime to another, by changing the nature of the main $p\lesssim\bar p$ terms and the nature of its $p\gg\bar p$ tail. The present result for the large adiabatic Holstein polaron, together with the well-known applicability of TPT to the Lang-Firsov limit for arbitrary $g$, suggests then strongly that TPT can cover the whole 2D parameter space of the Holstein Hamiltonian.

An interesting unsolved problem in which TPT can be useful refers to the evaluation of the polaron mass $M_p$ for the adiabatic polaron propagation through the PN barriers. Actually, the quantum tunneling through the PN barriers makes $M_p$ nonlinear in $M$ in a way not yet clearly distinguished from that in the nonadiabatic regimes.

Finally, it is clear that the problems, analogous to those opened and partially answered here, arise in connection to numerous quantum crossovers related to the symmetry breaking, and that some of the ideas developed here may apply to those cases too. As mentioned in Sec. \ref{SecIntroduction} this concerns in particular the problems of quantum solitons or doping the 2D Mott insulator, which are currently subjects of extensive theoretical investigation.

\begin{acknowledgments}

We wish to thank D. Feinberg, V. V. Kabanov, and A. S. Mishchenko for interesting correspondence. This work was supported by the Croatian Government under Projects No. $0035007$ and No. $0119256$. 

\end{acknowledgments}

\begin{widetext}
\appendix

\section{Appendix}

The crossing fourth-order diagram in Fig. \ref{Fig002}c is a convolution,

\begin{equation}
\Sigma_k^C(\omega)=
\frac{g^2}{N}\sum_qG_{k-q}(\omega-\omega_0)\;\Gamma_q(\omega)\;,
\label{App01}
\end{equation}

\noindent with $G_k(\omega)$ the free electron propagator and $\Gamma_q(\omega)$ the leading Migdal vertex correction, the latter given by a convolution

\begin{eqnarray*}
\Gamma_q(\omega)&=&\frac{g^2}{N}\sum_{q'}
G_{q'-q}(\omega-2\omega_0)\;G_{q'}(\omega-\omega_0)\\
&=&\frac{g^2}{t^2}\frac{1}{N}z_q\sum_{q'}\frac{z_{q'}^2}
{(z_{q'}-z_qy_+)(z_{q'}-z_qy_-)(z_{q'}-x_+)(z_{q'}-x_-)}\;,
\end{eqnarray*}

\noindent where $z_q=e^{iq}$ and 

\[\gamma_n=1+n\alpha-\omega/2t>1\;\;\;,\;
x_\pm=\gamma_1\pm\sqrt{\gamma_1^2-1}\;\;\;,\;
y_\pm=\gamma_2\pm\sqrt{\gamma_2^2-1}\;\;\;,\;
x_+x_-=y_+y_-=1\;.\]

\noindent As $x_-,y_-<1$, by integrating over the unit circle, $1/N\sum_q\rightarrow\oint d\phi/2\pi\rightarrow\oint_{|z|=1} dz/iz2\pi$, one obtains

\begin{eqnarray*}
\Gamma_q(\omega)&=&\frac{g^2}{t^2}z_q\left(
\frac{z_qy_-}{(z_qy_--z_qy_+)(z_qy_--x_+)(z_qy_--x_-)}
+\frac{x_-}{(x_--z_qy_+)(x_--z_qy_-)(x_--x_+)}\right)\\
&=&-\frac{g^2}{t^2}\frac{x_+y_+-x_-y_-}{(x_+-x_-)(y_+-y_-)}
\frac{z_q^{-1}}{(z_q^{-1}-\delta_+)(z_q^{-1}-\delta_-)}
\end{eqnarray*}

\noindent where $\delta_\pm=y_\pm/x_\mp$ is real, and $\delta_+\delta_-=1$. Substituting $\Gamma_q(\omega)$ into Eq.~(\ref{App01}) gives

\begin{eqnarray}
\Sigma_k^C(\omega)&=&-\frac{g^4}{t^3}z_k^{-1}
\frac{x_+y_+-x_-y_-}{(x_+-x_-)(y_+-y_-)}
\oint \frac{dz}{i2\pi}
\frac{1}{(z-z_k^{-1}x_+)(z-z_k^{-1}x_-)}\frac{z}{(z-\delta_+)(z-\delta_-)}
\nonumber\\
&=&-\frac{g^4}{t^3}\frac{x_+^2y_+-x_-^2y_-}{(x_+-x_-)^2(y_+-y_-)}
\frac{1}{x_+^2y_++x_-^2y_--2\cos{(k)}}\label{App02}
\end{eqnarray}

In the continuous limit one obtains:

\begin{eqnarray}
x_\pm&=&1+\varepsilon\pm\sqrt{2\varepsilon+\varepsilon^2}\approx
1+\varepsilon\pm\sqrt{2\varepsilon}\nonumber\\
y_\pm&\approx&1+\varepsilon+\alpha\pm\sqrt{2(\varepsilon+\alpha)}\nonumber\\
\Sigma_k^C(\omega)&=&-\frac{g^4}{(2t)^3}
\frac{2\sqrt{\varepsilon}+\sqrt{\varepsilon+\alpha}}
{\varepsilon\sqrt{\varepsilon+\alpha}}
\frac{1}{k^2+2(2\sqrt{\varepsilon}+\sqrt{\varepsilon+\alpha})^2}\;,\label{App03}
\end{eqnarray}

\noindent i.e., the result given by Eq. (\ref{Eq017}). The contribution to the effective polaron mass in Eq. (\ref{Eq007}) from $\Sigma_k^C(\omega)$ in the numerator is obtained as

\[t^{-1}\partial\Sigma_k^C(\omega)/\partial k^2|_{k=0}=\frac{1}{2}
\frac{g^4}{(2t)^4}\frac{1}{\varepsilon\sqrt{\varepsilon+\alpha}}
\frac{1}{(2\sqrt{\varepsilon}+\sqrt{\varepsilon+\alpha})^3}\sim\frac{g^4}{t^4}\frac{1}{\Lambda^6}
\sim\frac{t^2\omega_0^6}{g^8}\;.\]

\end{widetext}  


\begin{thebibliography}{99}

\bibitem{F} H. Fr\"ohlich,
	Proc. Roy. Soc. London, Ser. A {\bf 223}, 296 (1954).

\bibitem{Davydov} A. S. Davydov and N. I. Kislukha,
	Phys. Status Solidi B {\bf 59}, 465 (1973).

\bibitem{Gennes} P. -G. de Gennes,
	Phys. Rev. {\bf 118}, 141 (1960).

\bibitem{KT} J. M. Kosterlitz and D. J. Thouless,
	J. Phys. C {\bf 6}, 1181 (1973).

\bibitem{ZR} F. C. Zhang and T. M. Rice,
	Phys. Rev. B {\bf 37}, 3759 (1987).

\bibitem{L} L. D. Landau,
	Phys. Zeitschrift Sowjetunion {\bf 3}, 664 (1933).

\bibitem{LandauPekar} L. D. Landau and S. I. Pekar,
	Zh. Experim. i Teor. Fiz. {\bf 18}, 419 (1948).

\bibitem{Shaw} P. B. Shaw and E. W. Young,
	Phys. Rev. B {\bf 24}, 714 (1981).

\bibitem{Holstein4} T. D. Holstein and L. A. Turkevich,
	Phys. Rev. B {\bf 38}, 1901 (1988).

\bibitem{Gerlach} B. Gerlach and H. L\"{o}wen,
	Phys. Rev. B {\bf 35}, 4291 (1987);
	Rev. Mod. Phys. {\bf 63}, 63 (1991);
	H.~L\"{o}wen,
	Phys. Rev. B {\bf 37}, 8661 (1988).

\bibitem{A1} A. S. Alexandrov, V. V. Kabanov, and D. K. Ray,
	Phys. Rev. B {\bf 49}, 9915 (1994);
	A. S. Alexandrov and V. V. Kabanov,
	{\it ibid.} {\bf 54}, 3655 (1996).

\bibitem{A2} P. E. Spencer, J. H. Samson, P. E. Kornilovitch,
	and A. S. Alexandrov,
	Phys. Rev. B {\bf 71}, 184310 (2005).

\bibitem{A3} E. N. Myasnikov, A. E. Myasnikova, and F. V. Kusmartsev,
	Phys. Rev. B {\bf 72}, 224303 (2005).

\bibitem{A4} A. S. Alexandrov,
	Europhys. Lett. {\bf 56}, 92 (2001).

\bibitem{A5} J. P. Hague, P. E. Kornilovitch, A. S. Alexandrov, and J. H. Samson,
	Phys. Rev. {\bf 73}, 054303 (2006).

\bibitem{PN} R. E. Peierls,
	Proc. Phys. Soc. London {\bf 52}, 23 (1940);
	F. R. N. Nabarro,
	{\it ibid.} {\bf 59}, 256 (1947).

\bibitem{Dz} I. E. Dzyaloshinskii,
	Nobel Symposium {\bf 24}, 143 (1973) 
	(ed. B. Lundquist and S. Lundquist, Academic Press, N. Y. and London).

\bibitem{O} O. S. Bari\v si\' c,
	Phys. Rev. B {\bf 73}, 214304 (2006).

\bibitem{Rashba} E. I. Rashba,
	Opt. Spectrosk. {\bf 2}, 664 (1957).

\bibitem{H} T. Holstein,
	Ann. Phys. (N.Y.) {\bf 8}, 325 (1959).

\bibitem{AGD} A. A. Abrikosov, L. P. Gor'kov, and I. E. Dzyaloshinskii,
	{\it Methods of Quantum Field Theory in Statistical Physics}
	(Dover Publications, New York, 1963).

\bibitem{Fg} S. Ciuchi, F. de Pasquale, S. Fratini, and D. Feinberg,
	Phys. Rev. B {\bf 56}, 4494 (1997).

\bibitem{Mishchenko2} A. S. Mishchenko, N. Nagaosa, N. V. Prokof'ev,
	A. Sakamoto, and B. V. Svistunov,
	Phys. Rev. B {\bf 66}, R20301 (2002);
	A. S. Mishchenko and N. Nagaosa,
	Phys. Rev. Lett. {\bf 93}, 036402 (2004).

\bibitem{OSB} O. S. Bari\v si\' c and S. Bari\v si\' c,
	Fizika A {\bf 14}, 153 (2005),
	http://fizika.hfd.hr/fizika\_a/av05/a14p153.htm\ .

\bibitem{Mi} A. B. Migdal,
	Zh. Eksp. Teor. Fiz. {\bf 34}, 1438 (1958) [Sov. Phys. JETP {\bf 7}, 7989 (1958)].

\bibitem{Emin} D. Emin,
	Phys. Rev. B {\bf 48}, 13691 (1993);
	G. Kalosakas, S. Aubry, and G. P. Tsironis,
	{\it ibid.} {\bf 58}, 3094 (1998).

\bibitem{U} K. Uzelac and S. Bari\v si\' c,
	J. de Physique Letters {\bf 38}, L-47 (1977).

\bibitem{Kivshar} Y. S. Kivshar and D. K. Campbell,
	Phys. Rev. E {\bf 48}, R3077 (1993);
	D. Cai, A. R. Bishop, and N. Gr{\o}nbech-Jensen,
	{\it ibid.} {\bf 53}, 4131 (1996);
 	L. Brizhik, A. Eremko, L. Cruzeiro-Hansson, and Y. Olkhovska,
	Phys. Rev. B {\bf 61}, 1129 (2000).


\end{thebibliography}
\end{document}